# Astrophysical Observations: Lensing and Eclipsing Einstein's Theories

Charles L. Bennett

Albert Einstein postulated the equivalence of energy and mass, developed the theory of special relativity, explained the photoelectric effect, and described Brownian motion in five papers, all published in 1905, 100 years ago. With these papers, Einstein provided the framework for understanding modern astrophysical phenomena. Conversely, astrophysical observations provide one of the most effective means for testing Einstein's theories. Here, I review astrophysical advances precipitated by Einstein's insights, including gravitational redshifts, gravitational lensing, gravitational waves, the Lense-Thirring effect, and modern cosmology. A complete understanding of cosmology, from the earliest moments to the ultimate fate of the universe, will require developments in physics beyond Einstein, to a unified theory of gravity and quantum physics.

Einstein's 1905 theories form the basis for much of modern physics and astrophysics. In 1905, Einstein postulated the equivalence of mass and energy (*1*), which led Sir Arthur Eddington to propose (*2*) that stars shine by converting their mass to energy via $E = mc^2$, and later led to a detailed understanding of how stars convert mass to energy by nuclear burning (*3*, *4*). Einstein explained the photoelectric effect by showing that light quanta are packets of energy (*5*), and he received the 1921 Nobel Prize in physics for this work. With the photoelectric effect, astronomers determined that ultraviolet photons emitted by stars impinge on interstellar dust and overcome the work function of the grains to cause electrons to be ejected. The photoelectrons emitted by the dust grains excite the interstellar gas, including molecules with molecular sizes of ~1 nm, as estimated by Einstein in 1905 (*6*). Atoms and molecules emit spectral lines according to Einstein's quantum theory of radiation (*7*). The concepts of spontaneous and stimulated emission explain astrophysical masers and the 21-cm hydrogen line, which is observed in emission and absorption. The interstellar gas, which is heated by starlight, undergoes Brownian motion, as also derived by Einstein in 1905 (*8*).

Two of Einstein's five 1905 papers introduced relativity (*1*, *9*). By 1916, Einstein had generalized relativity from systems moving with a constant velocity (special relativity) to accelerating systems (general relativity).

Space beyond Earth provides a unique physics laboratory of extreme pressures and temperatures, high and low energies, weak and strong magnetic fields, and immense dimensions that cannot be reproduced in laboratories or under terrestrial conditions. The extreme astrophysical environments

Department of Physics and Astronomy, The Johns Hopkins University, 3400 North Charles Street, Baltimore, MD 21218, USA. E-mail: cbennett@jhu.edu

enable us to test Einstein's theories, and the general theory of relativity (10) has been tested and applied almost exclusively in the astrophysical arena.

### Gravity as Geometry

Newton formulated the first universal laws of gravity and motion. Using Newton's laws, astronomers can predict how moons will orbit planets. Newton, however, could not understand how a moon would sense the presence of a planet and wrote, "I have not been able to discover the cause of those properties of gravity from phenomena, and I frame no hypotheses…" (11). Einstein's theory of general relativity answered Newton's question: Mass causes space-time curvature. A planet bends the space around it, and its moon moves under the influence of that bent space.

Einstein encouraged astronomers to measure the positions of stars near the Sun to see how much the light was deflected. During an eclipse, it is possible to see stars near the occulted solar disk. An accurate measurement of the apparent position of one of these stars would provide a test of Einstein's idea (Fig. 1). Eddington traveled to Principe Island in western Africa for the 29 May 1919 eclipse, and he claimed to have seen evidence of the starlight shift predicted by Einstein. Today, we no longer are constrained to viewing optical light during eclipses; high-resolution radio interferometers, for example, can measure the position of quasar light that passes near the Sun. These recent measurements (12) provide more compelling support for relativity. Irwin Shapiro recognized that when light traverses curved space, such as near the Sun, it will arrive at its destination at a later time than light that passes through unperturbed flat space (13). Shapiro used radar ranging to planets and spacecraft to determine that the Sun's bending of space adds about 200 µs to the round-trip travel time of a radio signal passing near the edge of the Sun. The Shapiro time delay was recently confirmed by tracking the Cassini spacecraft during its flight to Saturn (14). The measurement agreed with relativity to 1 part in $10^5$.

Of the planets in the solar system, Mercury's orbit is most affected by the Sun's gravity (Fig. 1). The radius of Mercury's elliptical orbit (with an eccentricity of 0.206) varies from 46 to 70 million km. The measurement of the advance of the perihelion (the closest approach of an orbit to the Sun) of Mercury from the Earth is 5599 arc-sec per century, of which 5025 arc-sec per century are caused by the precession of Earth's equinoxes. The remaining 574 arc-sec per century do not match the Newtonian prediction that the axial direction of Mercury's elliptical orbit about the Sun should precess, advancing by 531 arc-sec per century, caused by perturbations from the other planets. In 1855, Le Verrier hypothesized the existence of an undetected planet, Vulcan, between Mercury and the Sun, to explain this discrepancy. Instead, general relativity accounts for the remaining 43 arc-sec per century and for the more constraining recent data acquired from radio tracking of the Mars landers.

During the 1930s, Fritz Zwicky determined that clusters of galaxies have 10 times

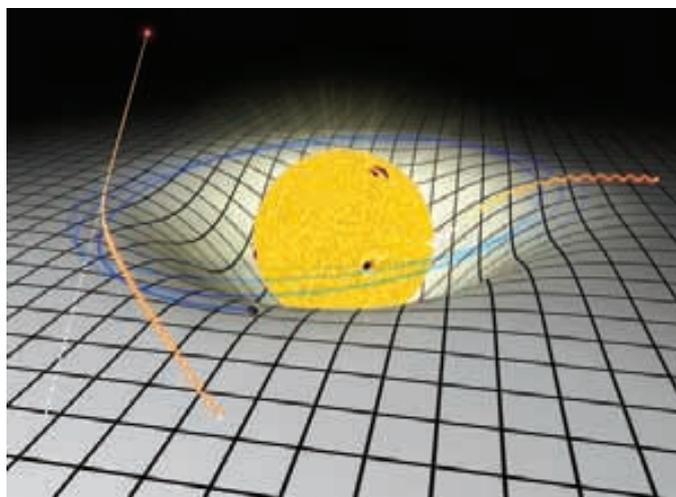

**Fig. 1.** A mass like our Sun bends space. On the right, light loses energy as it climbs the gravitational potential, undergoing a gravitational redshift. On the left, the path of distant starlight is deflected (or gravitationally lensed) by the curved space. Because of the relativistic effects of the curved space, the deflection angle is twice that predicted by Newtonian calculation. The advance of the perihelion of the orbit of the planet Mercury (seen in blue) is slightly altered by relativistic effects, as discussed in the text. [Figure by Theophilus Britt Griswold]

more mass than their light suggests, and termed the unseen mass "dark matter." He suggested that the matter would cause space to curve and act as a gravitational lens, and that the clusters could be "weighed" by the lensing effects (15, 16). The suggestion was not to test Einstein's definition of gravity, but to use it as a tool. Gravitationally lensed systems confirm Zwicky's assertion that clusters are dominated by dark matter (Fig. 2). Observations of the rotation curves of individual spiral galaxies also require large amounts of dark matter in halos around the galaxies (17).

### Gravitational Redshifts

A central tenet of relativity is Einstein's equivalence principle: It is impossible to distinguish between a local gravitational field and an equivalent uniform acceleration in a sufficiently small region of space-time. The gravitational redshift follows directly from this equivalence principle. Light that propagates through a gravitational field changes wavelength according to $\lambda/\lambda_0 = \Delta\Phi/c^2$, where $\lambda$ is the shifted wavelength, $\lambda_0$ is the rest wavelength, $c$ is the speed of light, and $\Delta\Phi$ is the change in the Newtonian gravitational potential (valid in the weak gravitational field limit). In strong gravitational fields that are a radial distance $r$ from a mass $M$ (where $r$ is defined as a circumference divided by $2\pi$), $\lambda/\lambda_0 = (1 - 2GM/c^2r)^{1/2}$, where $G$ is the gravitational constant. The observed wavelength goes to zero at the Schwarzschild radius, $R_S = 2GM/c^2$ (the event horizon). The gravitational redshift was confirmed to 10% accuracy in 1960 by using the transmission of a very narrow 14.4-keV transition from the isotope $^{57}$Fe across the 22.6-m height of the Jefferson Physical Laboratory building at Harvard (18). Later refinements of these Pound-Rebka-Snider experiments reached 1% accuracy. Vessot et al. (19) used a hydrogen maser clock on a rocket launched to an altitude of $10^4$ km to measure the gravitational redshift to an accuracy of $7 \times 10^{-5}$.

### Gravitational Waves

Maxwell's equations explain how an accelerating charged particle produces electromagnetic radiation. Analogously, general relativity predicts that an accelerating mass produces gravitational radiation. This radiation has not yet been measured directly, but it has been detected indirectly. From 1974 to 1983, Hulse and Taylor observed the pulsar (20) PSR1913+16, which orbits a 1.4–solar mass ($M_\odot$) star with a period of only 7.75 hours (21).

According to general relativity, these two massive bodies in rapid motion produce gravitational waves, and the emission of these waves should cause the binary system to lose energy. Indeed, the period of the pulsar's orbit was observed to be increasing at a rate of 76 millionths of a second per year, consistent (within 0.02%) with the expected energy loss caused by gravitational waves (22).

Considerable effort is now underway to make direct measurements of 10- to 1000-Hz gravitational radiation with the Laser Interferometer Gravitational Wave Observatory (LIGO). The Laser Interferometer Space Antenna (LISA), a planned three-spacecraft constellation, will measure gravitational waves in the range from 0.0001 to 1 Hz. LIGO and LISA (Fig. 3) will seek to detect gravitational waves from orbits, coalescences, and collisions of strong-field objects such as neutron stars and black holes. Their objective is to probe gravity in systems where the ratio of the gravitational binding energy to rest mass is in the range of 0.1 to 1, which is many orders of magnitude beyond any current measurement.

### The Lense-Thirring Effect

General relativity predicts the existence of black holes and compelling evidence has been found for their existence in many galaxies (23). Stars that exhaust their hydrogen supply cannot start burning carbon unless they have more than 4 $M_\odot$. Stars with less mass will collapse until the contraction is halted by electron degeneracy, according to the Pauli exclusion principle. These contracted stars are called white dwarfs. If the mass of the collapsing star is greater than 1.44 $M_\odot$, then electrons and protons will combine to form neutrons, and the star will collapse further until it is supported by the neutron degeneracy of the Pauli exclusion principle. These more massive contracted stars are called neutron stars. If the mass is greater than 2 to 3 $M_\odot$, even the neutron degeneracy cannot support the mass. With nothing to stop it, the star will collapse to become a black hole. The strong gravitational fields associated with neutron stars and black holes provide a laboratory for testing general relativity.

For a gyroscope near a rotating black hole, distant stars provide an inertial reference frame, but the black hole defines a different reference frame. The closer the gyroscope is to the black hole, the more it will be affected by the black hole's reference frame. The gyroscope will precess while trying to follow the rotation of the black hole, but it will lag behind because of its fidelity to the distant stars. In 1918, Joseph Lense and Hans Thirring introduced this problem, which does not arise in Newtonian gravity. The Lense-Thirring precession can be thought of as the dragging of inertial frames around rotating masses. The rotation twists the surrounding space, perturbing the orbits of nearby masses.

Possible x-ray detections of frame dragging around neutron stars and black holes have been suggested (24). Unfortunately, attempts to measure the Lense-Thirring effect around black holes are often confounded by the complexities of the dynamics of the hot gas in their accretion disks. Earth's gravitational field, although much weaker, should also cause frame dragging. A detection of the Lense-Thirring effect has been claimed with 10% accuracy based on the precession of the orbital planes of the two LAser GEOdynamics Satellites (LAGEOS) (25). LAGEOS-1 and LAGEOS-2 are 0.6-m-diameter aluminum spherical shells in Earth orbit, each covered with 426 passive retroreflectors.

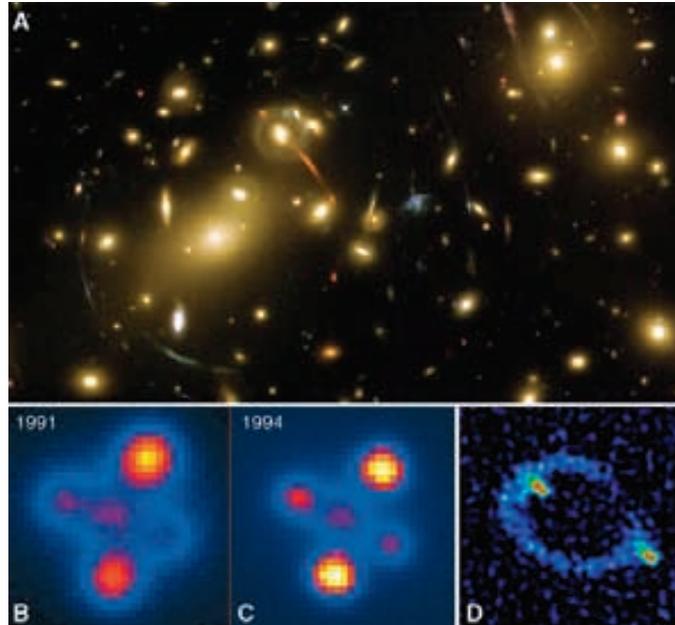

**Fig. 2.** The distribution of dark matter in space can be deduced from gravitational lensing effects. (**A**) An HST picture of the galaxy cluster Abell 2218 shows distorted images of the background light, including arcs, contorted galaxy shapes, and multiple images of single objects. [Courtesy of NASA by A. Fruchter and the HST Early Release Observations Team] With sources of variable luminosity, differential propagation time delays become apparent in the multiple images, such as with the 3-year change in the Einstein cross (**B** and **C**). In a coaligned system (source to lens, mass to observer), the source appears as an Einstein ring (**D**). [Courtesy of G. Lewis (Institute of Astronomy) and M. Irwin (Royal Greenwich Observatory) from the William Hershel Telescope and J. N. Hewitt and E. L. Turner from the National Radio Astronomy Observatory/Associated Universities, Inc.]

Analogous to an accelerating charge producing a magnetic field, an accelerating mass produces certain ''gravitomagnetic'' effects. Three gravitomagnetic effects are: (i) the precession of a gyroscope in orbit about a rotating mass, (ii) the precession of orbital planes in which a mass orbiting a large rotating body constitutes a gyroscopic system whose orbital axis will precess, and (iii) the precession of the pericenter of the orbit of a test mass about a massive rotating object. NASA's Gravity Probe B (GPB) seeks to detect and measure the spin-spin interaction between its own orbiting gyroscope and the spinning Earth (frame dragging) and the combination of the effects of spin-orbit coupling and space-time curvature known as geodetic (or Thomas) precession.

### Big Bang Nucleosynthesis

Big Bang nucleosynthesis calculations depend on the relation between temperature and the rate of expansion of the universe. This relation is specified by the Friedmann equations, which in turn are derived from Einstein's field equations. Thus, the degree to which Big Bang nucleosynthesis matches the measured abundance ratios of the light chemical elements is also a test of Einstein's equations. In 1946, George Gamow wrote that it was ''generally agreed … that the relative abundance of various chemical elements were determined by physical conditions existing in the universe during the early stages of its expansion, when the temperature and density were sufficiently high to secure appreciable reaction-rates for the light as well as for the heavy elements.'' (26). Gamow gave the problem to his graduate student, Ralph Alpher, to work out. Alpher's Ph.D. thesis and a resulting journal paper (27) appeared in 1948, showing in detail that the lightest elements (with atomic numbers ≤4) would be formed in the early universe. Gamow was wrong about the heavier elements, which we now know were created later in stellar processes (28).

### Our Dynamic Universe

General relativity is at the heart of modern cosmological models. Einstein's field equation is $R_{\mu\nu} - \tfrac{1}{2}Rg_{\mu\nu} = 8\pi GT_{\mu\nu}$, where $R_{\mu\nu}$ is the Ricci tensor, $R$ is the Ricci scalar, $G$ is Newton's gravitational constant, $g_{\mu\nu}$ is the metric tensor, and $T_{\mu\nu}$ is the stress-energy tensor. The left-hand side of the equation describes the curvature of space and the right-hand side describes the content of space. Thus, the equation reflects the equivalence of matter and curved space. When Einstein applied his field equation to cosmology, he found a problem. His equations implied an unstable (evolving) universe. To produce a stable universe, he needed to add a parameter to his equation in order to overcome the attractive gravity of all of the mass. In 1917, he introduced what he called the cosmological constant as a repulsive antigravity term into the field equation. The field equation with Einstein's cosmological con-stant term, $\Lambda$, becomes $R_{\mu\nu} - \frac{1}{2}Rg_{\mu\nu} + \Lambda g_{\mu\nu} = 8\pi GT_{\mu\nu}$. The cosmological constant can be thought of as the energy of the vacuum (that is, how much "'nothing'' weighs) and is a constant of nature, with no specified value. Einstein fine-tuned its value to balance exactly the tendency of mass in the universe to collapse gravitationally.

Now Einstein had a static finite universe, but with a second fundamental gravitational constant of nature. Not only was this a fine-tuned solution, but it was unstable to perturbations; the tiniest of fluctuations would cause the universe to shrink or to expand.

This solution was also based on the erroneous assumption that the universe is static. In 1929, Edwin Hubble determined that the recession velocities of galaxies (determined by Doppler shifts) are proportional to their distances. Assuming that we do not occupy a special place in the universe, Hubble concluded that the universe is expanding. Given an expanding universe and not a static one, Einstein concluded that his cosmological constant was unnecessary.

For Hubble, Cepheid variable stars provided the key connection to obtain the distance to galaxies. The recession velocities of galaxies were determined by the Doppler shift of their light. More recently, Type Ia supernovae light curves have been used for distance determinations. In an attempt to measure the slowing of the expansion of the universe by using supernovae, astronomers discovered that the expansion of the universe is accelerating, not decelerating (*29*, *30*). Einstein's cosmological constant has reemerged as one of many candidate explanations for the accelerated expansion.

### The Cosmic Microwave Background

In working through Big Bang nucleosynthesis in 1949, Alpher and Herman estimated (*31*) that blackbody radiation with "a temperature on the order of 5 K" would remain today, "interpreted as the background temperature which would result from the universal expansion alone." Arno Penzias and Robert Wilson won the 1978 Nobel Prize for their 1965 detection of this afterglow of the Big Bang. The radiation has now been precisely and accurately characterized by using the Cosmic Background Explorer (COBE) satellite as a black body with a radiation temperature of $2.725 \pm 0.001$ K (Fig. 4A) (*32*). The blackbody spectrum provided evidence that the radiation had been in thermal equilibrium with matter earlier in the history of the universe, as predicted within the Big Bang model.

Penzias and Wilson found the cosmic microwave background (CMB) to be uniform across the sky to within the precision of their measurements. Further measurements were performed in order to detect the nonuniformity that was expected because the gravitational seeds of the clumpy structures we see in the sky today should be associated with small temperature variations across the sky in the CMB. Year after year, experimentalists improved technology to reach ever fainter upper limits on the amplitude of these temperature fluctuations, and the lack of detectable fluctuations began to be a problem. Without large enough primordial gravitational fluctuations, there wouldn't be sufficient time for gravity to cause the growth of the current structure of the universe. A solution to the structure problem came with the introduction of cold dark matter (CDM) (*33*). Although baryons may not always emit much light and may be thought of as dark matter, CDM is fundamentally different because CDM particles do not interact with light.

In the early universe, when it was too hot for electrons to bind to hydrogen atoms, there was a plasma of photons, electrons, baryons, and CDM. The photons would frequently scatter off of the free electrons. The seeds of the structure of the universe were also present in the form of primordial gravitational perturbations. The CDM would congregate in regions of gravitational potential minima. The combined gravitational pull of the primordial fluctuation and the CDM would attract the baryonic matter, but unlike the CDM, the baryonic matter would experience enormous radiation pressure that would prevent the baryons from clumping.

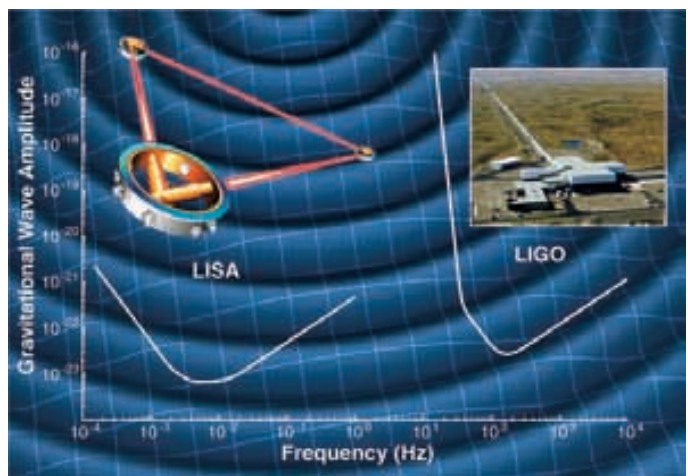

**Fig. 3.** LIGO is searching for gravitational waves of 3 to 100 ms, which are expected from the collapsing cores of supernovae and from coalescing neutron stars and stellar-mass black holes. LISA is sensitive to gravitational waves with much longer periods and is currently under development, with the goal of detecting events such as the coalescence of supermassive black holes.

The opposing forces of gravitational attraction and radiation pressure induced adiabatic oscillations in the photon-baryon fluid.

This occurred within the general relativistic expansion and cooling of the universe. As the universe expanded, the radiation pressure dropped and the universe became matter-dominated. At that point, the baryons also participated in the growth of structures. When the universe cooled sufficiently for electrons to bind to protons to form neutral hydrogen ($\sim$3000 K), the photons no longer scattered. These CMB photons we now observe carry with them the signature of the epoch of last scattering, called decoupling. On an angular scale greater than 2°, the photons have been gravitationally redshifted according to the amplitude of the gravitational potential at decoupling.

In 1992, the 27-year search for the anisotropy in the CMB ended with the COBE discovery of the faint fluctuations in a map of the full sky (Fig. 4B). This discovery supported the basic cosmological picture that structure developed gravitationally in the universe, with the mass dominated by CDM.

The detection of the primordial gravitational anisotropy caused experimentalists to redirect their efforts to the characterization of the statistics of the anisotropy, especially at smaller angular scales. A precise and accurate measurement of the statistics of the anisotropy pattern was predicted to reveal the signature of the adiabatic oscillations, the nature of which depends on the values of key cosmological parameters. CMB balloon-borne (*34*) and ground-based (*35*) experiments collected a wealth of data (*36*) on the nature of the adiabatic oscillations, and the space-borne Wilkinson Microwave Anisotropy Probe (WMAP) mission produced full-sky multi-frequency maps of the CMB (Fig. 4C) that tightly constrain the dynamics, content, and shape of the universe (*37*).

Observational cosmologists have been productive on many other fronts in the past decade. The Hubble Space Telescope (HST) determined the expansion rate of the universe, known as the Hubble constant (*38*). The 2° Field Galaxy Redshift Survey and the Sloan Digital Sky Survey have constrained cosmological parameters (*39*, *40*) by determining the three-dimensional power spectrum based on observations of hundreds of thousands of galaxies.

The combination of the cosmological observations of the past 10 years has led to a highly constrained cosmological model that is consistent with these wide-ranging observations and general relativity: A flat (Euclid-ean) universe is dominated by 73% dark energy, 23% CDM, and 4% baryons, and has a Hubble constant of 71 km s$^{-1}$ Mpc$^{-1}$. Despite this knowledge, we still do not know the specific nature of the dark energy or the CDM, nor do we understand the physics of the origin of the universe.

### Inflation

There is evidence that the universe underwent a short early burst of extremely rapid expansion, called inflation. Inflation has been used to explain the so-called flatness, horizon, and monopole problems. We measure space to be nearly flat because the inflationary expansion drove the initial radius of curvature of the universe to become large. We find the CMB temperature to be highly uniform across the sky because, with inflation, regions more than 2° apart on the sky were in causal contact and thus had an opportunity to equilibrate. We observe no magnetic monopoles because the inflationary expansion drove their density to an infinitesimal level.

Inflation is a paradigm, and within the inflationary paradigm there are many specific models. Each model is characterized by a tensor-to-scalar ratio (that is, the ratio of anisotropy power in gravitational waves to adiabatic density perturbations) and a spectral index of spatial fluctuations, both of which are measurable quantities from CMB observations. Inflation generically produces gravitational waves. Inflationary gravitational waves will be exceedingly difficult to detect directly; however, they imprint a signature polarization pattern onto the CMB. The measurement of this pattern could determine the energy level of the inflationary expansion. The energy level of inflation is likely to be at the scale of the grand unified theory (GUT), where the strong and electroweak forces are unified. For GUT-scale inflation, the detection of the polarization patterns should be possible, albeit very difficult. CMB researchers have already begun the search for the primordial gravitational waves. A future NASA mission, the Einstein Inflation Probe, is currently in its early planning stages and is aimed at detecting the primordial gravitational waves by their imprint on the CMB.

### Unification

James Clerk Maxwell is famous for presenting the unified theory of electricity and magnetism called electromagnetism. The electromagnetic force was later quantized by Feynman in a theory called quantum electrodynamics (QED). The quantized electromagnetic force was then unified with the weak force by Weinberg, Salam, and Glashow. Analagous to the interactions between photons and charged particles in QED, quarks carry a special kind of "strong charge," arbitrarily called "color," and interact by the strong force

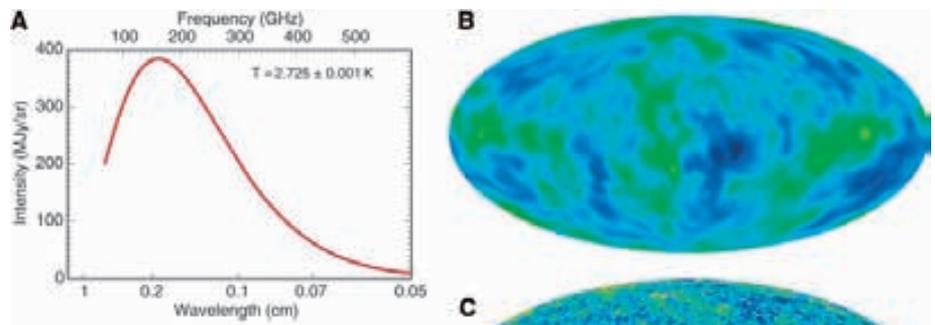

**Fig. 4.** NASA's COBE space mission made a precision measurement of the spectrum of the CMB radiation. The measurement uncertainties are smaller than the thickness of the red curve in the plot (**A**). COBE also made the first detection of the primordial anisotropy of the cosmic radiation, seen in the full-sky temperature map (**B**), where yellow is warmer and blue is cooler. (A model of foreground microwave emission and the Doppler term have been subtracted.) The primary physical effect is the gravitational redshifting of the light from primordial gravitational potential wells and blueshifting from primordial gravitational potential peaks. (**C**) The higher sensitivity and higher resolution WMAP full-sky map reveals the interactions of the photon-baryon fluid in the early universe, allowing a host of cosmological parameters to be deduced. [Photo courtesy of NASA's COBE and WMAP science teams]

in the theory of quantum chromodynamics (QCD). Gross, Wilczek, and Politzer recently shared the 2004 Nobel Prize in physics for their work in QCD. Because the electromagnetic force and the gravitational force both follow $1/r^2$ force laws (the force is inversely proportional to the square of the radial distance $r$), Einstein felt that these were the most promising forces to try to unify, yet he was unsuccessful. In 1919, Kaluza proposed to unify electromagnetism with gravity by introducing an extra dimension, and in 1926, Klein proposed that the extra dimension be compact. Some string theories predict that spacetime has 10 dimensions. If six dimensions are tightly curled up (compacted), then it would be difficult to establish their existence. We have now eclipsed Einstein with ongoing progress toward a unified quantum field theory and with new ideas for the unification of gravity from superstring and M theory.

### Dark Energy

We now believe that the universe is dominated by a completely unpredicted dark energy. Einstein's cosmological constant could be the dark energy, and all cosmological measurements are consistent with this interpretation. In this scenario, the energy densities of radiation and matter decrease as the universe expands, but the cosmological constant does not. Although the cosmological constant was negligible in the early universe, it now dominates the energy density of the universe (*41*). However, the cosmological constant as a dark energy candidate is unsatisfying because we cannot explain why its observed density is such that it just recently dominates the dynamics of the universe. It is tempting to associate the dark energy with the energy of the vacuum. Physical scaling suggests that the vacuum energy should be enormous, although the cosmological constant is tiny by comparison. Either there is a subtle reason for a near-zero but finite vacuum energy, or the dark energy is not vacuum energy.

The dark energy could be a sign of the breakdown of general relativity. A more general gravity theory may solve the problem, and research continues into various modified gravity theories. Alternatively, the dark energy could be a universally evolving scalar field (*42*).

Many other explanations for the dark energy have also been posited. Although it is often said that we have no idea what the dark energy is, it may be more correct to say that we have too many ideas of what the dark energy might be. And, of course, all of our current ideas may be wrong.

Progress in dark energy research is a high-priority challenge for theoretical and experimental physicists. Currently, two leading approaches for characterizing the dark energy are the use of supernovae light curves as distance indicators and the study of weak gravitational lensing. Weak lensing takes advantage of the contortions of the shapes of distant objects by intervening objects (lenses). Large statistical lensing samples, as a function of redshift, probe both the geometry and dynamics of the universe.

Einstein launched a revolution in physics that continues to this day, and his theories have undergone increasingly rigorous tests. Effects predicted by Einstein's theories have been used as tools to comprehend the universe. In many ways, we are now going beyond Einstein (*43*). On the theoretical frontier, efforts toward the unification of the laws of physics continue. On the experimental frontier, tireless efforts are underway to detect CDM directly in underground low-background experiments; to produce CDM in particle accelerators; to make measurements to determine the nature of the dark energy (its sound speed, equation of state, and evolution); and to test inflation and determine what happened in the earliest moments of our universe.